 \documentclass[final,3p,authoryear]{elsarticle}
\usepackage{amssymb}
\usepackage{lipsum}
\usepackage{amsmath,amssymb,amsfonts}
\usepackage{algorithmic}
\usepackage{graphicx}
\usepackage{textcomp}
\usepackage{xcolor}
\usepackage[inkscapelatex=false]{svg}
\usepackage{multirow}
\usepackage{makecell}
\usepackage{booktabs}
\usepackage{algorithm}
\usepackage{algorithmic}
\usepackage[caption=false,font=normalsize,labelfont=sf,textfont=sf]{subfig}
\usepackage{subcaption}

\usepackage{amssymb}
\usepackage{amsmath}
\journal{Nuclear Physics B}

\begin{document}

\begin{frontmatter}



\title{FedPCL-CDR: A Federated Prototype-based Contrastive Learning Framework for Privacy-Preserving Cross-domain Recommendation}

\author[first]{Li Wang}
\ead{li.wang-13@student.uts.edu.au}
\author[first]{Qiang\ Wu}
\ead{Qiang.Wu@uts.edu.au}
\author[first]{Min Xu\corref{cor1}}
\ead{Min.Xu@uts.edu.au}
\affiliation[first]{organization={School of Electrical and Data Engineering, University of Technology Sydney},%Department and Organization
            addressline={15, Broadway, Ultimo}, 
            city={Sydney},
            postcode={2000}, 
            state={NSW},
            country={Australia}}
\cortext[cor1]{Corresponding author}

%% Abstract
\begin{abstract}
Cross-domain recommendation (CDR) aims to improve recommendation accuracy in sparse domains by transferring knowledge from data-rich domains. However, existing CDR approaches often assume that user-item interaction data across domains is publicly available, neglecting user privacy concerns. Additionally, they experience performance degradation with sparse overlapping users due to their reliance on a large number of fully shared users for knowledge transfer. To address these challenges, we propose a Federated Prototype-based Contrastive Learning (CL) framework for Privacy-Preserving CDR, called FedPCL-CDR. This approach utilizes non-overlapping user information and differential prototypes to improve model performance within a federated learning framework. FedPCL-CDR comprises two key modules: local domain (client) learning and global server aggregation. In the local domain,  FedPCL-CDR first clusters all user data and utilizes local differential privacy (LDP) to learn differential prototypes, effectively utilizing non-overlapping user information and protecting user privacy. It then conducts knowledge transfer by employing both local and global prototypes returned from the server in a CL manner. Meanwhile, the global server aggregates differential prototypes sent from local domains to learn both local and global prototypes. Extensive experiments on four CDR tasks across Amazon and Douban datasets demonstrate that FedPCL-CDR surpasses SOTA baselines. We release our code at https://github.com/Lili1013/FedPCL\_CDR.
\end{abstract}


% \begin{highlights}
% \end{highlights}

%% Keywords

\end{frontmatter}




%% Use \subsection commands to start a subsection.
\section{Introduction}
\label{introduction}

Cross-domain recommendation (CDR) offers an effective solution to the data sparsity problem in recommendation systems by enabling knowledge transfer across different domains \citep{hu2018conet}. Based on different recommendation scenarios, existing CDR can be divided into two categories: single-target CDR and multi-target CDR. The first genre \citep{zhu2019dtcdr,zhu2022personalized} aims to improve recommendation performance in the target domain by utilizing rich information from the source domain. However, it cannot enhance model performance across multiple domains simultaneously. To address this issue, multi-target CDR \citep{9514447,zhu2023domain} has emerged, where each domain can act as either a source domain or a target domain. It leverages data from multiple domains to simultaneously enhance recommendation performance across all domains. These two kinds of methods apply some technologies, such as feature aggregation \citep{zhu2019dtcdr,zhu2020graphical} and feature disentanglement \citep{cao2022disencdr,guo2023disentangled}, to improve recommendation performance. Feature aggregation methods typically learn representations in each domain separately and then design an aggregation function to combine these representations. On the other hand, feature disentanglement approaches concentrate on separating domain-invariant and domain-specific representations and transferring domain-invariant representations to other domains. Despite their promising performance, these methods still face two major challenges.

\textbf{CH1. How to effectively protect user privacy when transferring cross-domain knowledge?}
Most existing methods \citep{zhu2020graphical,xu2023neural,lu2023contrastive} assume direct transfer of user-item ratings or representations across domains, making them unsuitable for privacy-preserving settings \citep{tian2024privacypreserving}. To address this, privacy-preserving CDR (PPCDR) methods \citep{chen2022differential,chen2023win,liu2024reducing} have gained lots of attention. For instance, PriCDR \citep{chen2022differential} uses Differential Privacy (DP) to release the source rating matrix for CDR modeling, while P2FCDR \citep{chen2023win} applies Local Differential Privacy (LDP) and Federated Learning (FL) to protect user embeddings. However, these methods  provide limited privacy protection, as attackers can exploit external information to infer sensitive data. For example, attackers can exploit a public dataset to learn reference embeddings and correlate them with differentially private embeddings to infer user preferences.

\begin{figure}[!t] %H为当前位置，!htb为忽略美学标准，htbp为浮动图形
\setlength{\abovecaptionskip}{0.1cm}
\centering %图片居中
\includegraphics[width=0.6\textwidth]{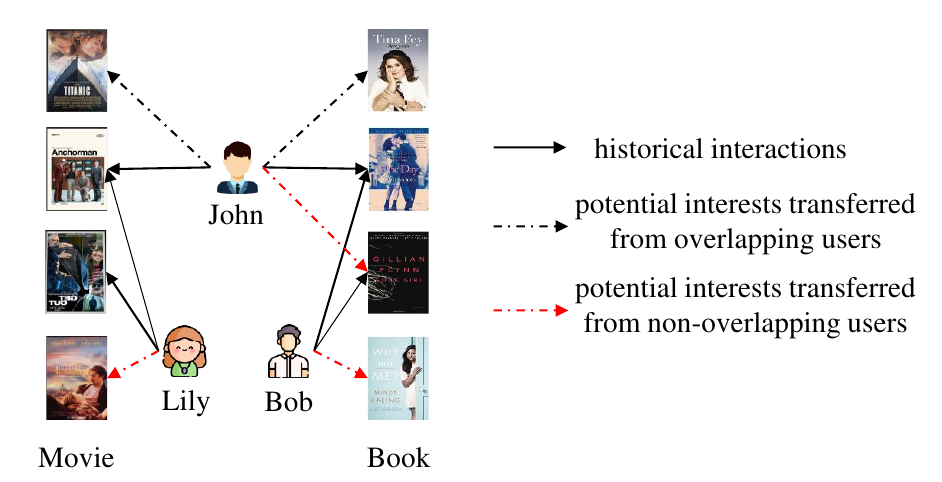} %插入图片，[]中设置图片大小，{}中是图片文件名
\caption{The illustrations of knowledge transfer via overlapping and non-overlapping users.}
\label{toy_example} %用于文内引用的标签
%\vspace{-1mm}
\end{figure}

\textbf{CH2. How to improve recommendation performance with sparse overlapping users across domains?} Many existing CDR methods depend on fully overlapping users as a bridge to transfer cross-domain knowledge \citep{cao2022disencdr,zhu2020graphical}. For example, as shown in Figure \ref{toy_example}, John is an overlapping user with interactions in both the Movie and Book domains. If knowledge transfer is solely based on fully overlapping users, John may like the comedic book ``Bossypants" because he watches movies in the same genre, like ``Anchorman."  However, in real-world datasets, there are very few overlapping users. For instance, the Amazon dataset has only a 5\% overlapping user ratio \citep{lu2023contrastive}. As illustrated in Figure \ref {toy_example}, non-ovelapping users like Lily and Bob have interactions in only one domain, making it impossible to directly transfer their preferences learned from other domains. As a result, the model performance will be degraded with such sparse overlap.

To address these challenges, we propose FedPCL-CDR, a federated prototype-based contrastive learning (CL) framework for PPCDR. It consists of two key modules, i.e., local domain learning and global server aggregation, where user-item interaction histories and review texts are stored in local domains and knowledge is transferred via differential prototypes within the FL framework, ensuring user privacy. Specifically, in the local domain, we first utilize k-means clustering \citep{macqueenclassification} to generate prototypes. On one hand, these prototypes convey overlapping and non-overlapping user preferences as they are derived from the interest alignments of all entities in the domain. Based on these prototypes, non-overlapping user preferences can be transferred across domains to improve recommendation performance. For example, in Figure \ref {toy_example}, Lily shares similar interests with John based on their common interaction with the movie ``Anchorman". By effectively utilizing Lily's interests, we can infer that John may like the thriller book ``Gone Girl" based on Lily's preference for the thriller movie ``Get Out", which couldn't be realized by relying solely on fully overlapping users. On the other hand, these prototypes serve as generalized representations of group-level preferences, making it more difficult for attackers to infer sensitive information about individual users. To further enhance privacy, we employ LDP technology to learn differential prototypes, which are then uploaded to the global server. The global server, in turn, models both local and global prototypes by aggregating these differential prototypes and sends them back to the respective local domains. To effectively transfer knowledge, the local domain refines user embeddings by using both local and global prototypes in a CL manner. This dual-prototype approach allows for transferring knowledge at varying granularities, enabling more nuanced learning of user embeddings from different perspectives.

To summarize, we highlight four main contributions of this work: 
\begin{itemize}
    \item We propose a novel federated prototype-based CL framework for PPCDR that aims to address the sparse overlapping user and privacy protection concerns.
    \item We explicitly leverage non-overlapping users for cross-domain knowledge transfer via clustering-based prototypes. This addresses the critical limitation of sparse overlapping users in real-world scenarios. In addition, we introduce a dual-prototype contrastive mechanism (local and global prototypes) to align user embeddings across domains at varying granularities, which captures finer-grained semantic relationships, enabling more robust knowledge transfer. 
    \item We utilize differential prototypes to transfer cross-domain knowledge within the FL framework, achieving stronger privacy protection.
    \item Extensive experiments on four CDR tasks, conducted on Amazon and Douban datasets, show that FedPCL-CDR outperforms SOTA baselines.
\end{itemize}

\section{Related Work}
In this section, we provide a brief overview of related research work, including cross-domain recommendation, privacy-preserving CDR, and contrastive learning.
\subsection{Cross Domain Recommendation}
Cross-domain recommendation (CDR) focuses on addressing the data sparsity problem by transferring knowledge across diverse domains or platforms. A core challenge in CDR is designing effective knowledge transfer methods from data-rich to data-sparse domains to improve recommendation accuracy. Recent deep learning approaches have explored various transfer techniques, including 
%domain transfer techniques \citep{yu2020semi}, 
domain mapping \citep{man2017cross}, feature combination \citep{zhu2020graphical}, feature alignment \citep{zhang2023disentangled},
%deep dual knowledge transfer \citep{li2020ddtcdr}, 
and graph-based methods \citep{cui2020herograph, zhao2019cross}.
%Our method is related to the third genre, which utilizes CL to realize feature alignment between user embeddings and prototypes. While various transfer methods make contributions to the performance improvement, 
For example, EMCDR \citep{man2017cross} introduces a MLP network to capture the non-linear mapping relationships across domains. GA-DTCDR \citep{zhu2020graphical} first learns user and item embeddings through separate heterogeneous graphs within the domain and then combines these cross-domain embeddings using an attention network. DCCDR \citep{zhang2023disentangled} is designed to disentangle domain-invariant and domain-specific features while aligning the domain-invariant features across domains. HeroGRAPH \citep{cui2020herograph} proposes a heterogeneous graph framework that unifies multiple domains into a shared graph structure, enabling cross-domain recommendation through meta-path-based neighbor aggregation and relation-aware attention mechanisms.
However, these approaches assume open cross-domain data sharing, neglecting privacy concerns. In addition, they rely on fully overlapping users or items to connect different domains and transfer knowledge, leading to performance degradation with sparse overlapping entities.
In this paper, we leverage non-overlapping user information and differential prototypes within the FL framework to address the sparse overlapping problem and mitigate the risk of user privacy leakage.
% \vspace{-0.5cm}
% \subsection{Federated Learning}
% Federated learning (FL) is a distributed machine learning approach that enables training models across multiple devices or servers while keeping data localized \citep{zhang2021survey}. This paradigm has been applied in various domains, such as healthcare \citep{brisimi2018federated}, natural language processing \citep{mcmahan2017communication} and recommendation systems \citep{perifanis2022federated,chai2020secure, zhang2023dual}, to address privacy and data security concerns. FL-based recommendation methods protect user privacy by training models locally on user devices and sharing only aggregated model parameters, rather than raw user data.  For example, FedNCF \citep{perifanis2022federated} is a federated collaborative filtering model that enables decentralized training while maintaining user privacy. Similarly, FedMF \citep{chai2020secure} extends matrix factorization techniques to the federated setting, allowing for private and efficient collaborative filtering. Another notable approach is FedCDR \citep{yan2022fedcdr}, which leverages the FL framework to hinder the leakage of user privacy when transferring knowledge across domains. 
% However, these FL-based methods focus on single-domain or single-target CDR scenarios and may not be directly applicable to multi-target CDR settings.
% % \vspace{-0.5cm}

\subsection{Privacy-preserving CDR}
With the enactment of privacy protection laws and the increasing focus on user privacy, many scholars have begun studying PPCDR methods. These approaches leverage technologies, such as LDP \citep{chen2022differential,chen2023winwin,gao2019privacypreserving, wang2024privacypreserving} and FL \citep{tian2024privacypreserving,liu2023federateda,yan2022fedcdr}, to protect users' sensitive information when transferring knowledge across domains. For example, PriCDR \citep{chen2022differential} employs Differential Privacy (DP) to release the source domain's rating matrix. P2FCDR \citep{chen2023winwin} utilizes the LDP technique to add noise to the transformed embeddings before transfer. PPGenCDR \citep{liao2023ppgencdr} further leverages adversarial methods to generate fake ratings for transfer to the target domain. FPPDM \citep{liu2023federateda} and FedCDR \citep{yan2022fedcdr} introduce the FL framework to protect user privacy through distributed learning. $P^2$DTR \citep{lin2024enhancing} introduces a dual-target CDR framework that utilizes a private set intersection algorithm and prototype-based FL to protect user privacy. Among these, $P^2$DTR is the most closely related to our work. However, our method introduces critical advancements that address key limitations: (1) $P^2$DTR primarily relies on overlapping users to learn prototypes and transfer knowledge, which limits its effectiveness in sparse-overlap scenarios. In contrast, we explicitly leverage non-overlapping users by clustering all users (both overlapping and non-overlapping) to extract generalized prototypes. This allows knowledge transfer even in scenarios with extremely limited overlapping users, significantly improving robustness in real-world sparse settings; (2) $P^2$DTR transfer cross-domain knowledge through Constrained Dominant Set (CDS) propagation, which primarily optimizes intra-domain user embeddings using averaged prototypes. Our work, however, introduces prototype-based CL, aligning user embeddings with both local prototypes and global prototypes. This dual-prototype strategy captures finer-grained semantic relationships, improving transfer effectiveness.

% Although these PPCDR methods have achieved significant success in protecting privacy, they still exhibit limited privacy-preserving ability and suboptimal model performance. In this paper, we propose the FUPM framework to address these challenges.

\subsection{Contrastive Learning}
Contrastive Learning (CL) has been widely used in computer vision \citep{chen2020simple,he2020momentum} and natural language processing \citep{gao2021simcse,chuang2022diffcse}.
It is a self-supervised learning technique that aims to maximize the mutual information between two representations. To achieve this, InfoNCE \citep{oord2018representation} is proposed to 
learn representations by contrasting positive pairs (similar samples) against negative pairs (dissimilar samples), which
discovers the semantic information shared by different views. Nowadays, CL has been applied to the recommendation field to improve representation learning \citep{wu2021self,zhang2023disentangled,chen2023heterogeneous,lu2023contrastive}. For instance, DCCDR \citep{zhang2023disentangled} leverages CL to learn domain-specific and domain-invariant representations. Meanwhile, CL-DTCDR \citep{lu2023contrastive} utilizes CL to learn more representative user and item embeddings with user-item interaction data and side information. 
However, these methods directly utilize user-item ratings or representations to construct positive and negative pairs across domains, which is not feasible under privacy-preserving constraints.
In this work, we employ differential prototypes to transfer user interests in a CL manner, thereby protecting user privacy.
% \vspace{-0.5cm}
% \subsection{Prototype Learning}
% Prototype learning has been used in all kinds of tasks, i.e., transfer learning \citep{quattoni2008transfer}, few-shot learning \citep{snell2017prototypical}, and interpretable machine learning \citep{bien2011prototype}. It represents the mean vector of the samples within a class or the centroid of a cluster. As prototypes can generalize semantic knowledge from similar samples, they play an important role in aiding local training within federated learning setups. FedProto \citep{tan2022fedproto} considers class prototypes as a bridge connecting local clients and the global server in the framework of federated learning. Authors in \citep{michieli2021prototype} present a novel approach to federated learning for visual data by leveraging prototype learning to improve feature representation, communication efficiency, and model interpretability. 
% Inspired by these methodologies, we integrate prototypes and federated learning into CDR, enabling knowledge transfer without disclosing user-item interaction information. 

\section{Methodology}
\subsection{Definitions and Notations}  We assume there are $M$ domains (clients) and a global server, where $D^i$ denotes the i-th domain. 
Within each domain, there exists a user set $U^i$ and an item set $V^i$. %with $m_i$ users and $n_i$ items. 
There are partial overlapping users, denoted as $U^o$. 
Let $\textbf{R}^i\in \{0,1\}^{|U^i|\times |V^i|}$ represent the binary user-item interaction matrix. 
% In addition, except for the user-item interaction matrix $\textbf{R}^i$, we don't introduce other auxiliary information, which is more common in real-world applications.
%The goal of our method is to recommend top-N items for all users in each domain. 
%Figure \ref{framework} depicts the overall framework of FedPCL-CDR. 
We illustrate the paradigm for domain $D^i$, and the corresponding paradigm for other domains can be easily inferred accordingly.

\subsection{Overall Framework}
\begin{figure*}[!t] %H为当前位置，!htb为忽略美学标准，htbp为浮动图形
\centering %图片居中
\includegraphics[width=1\textwidth]{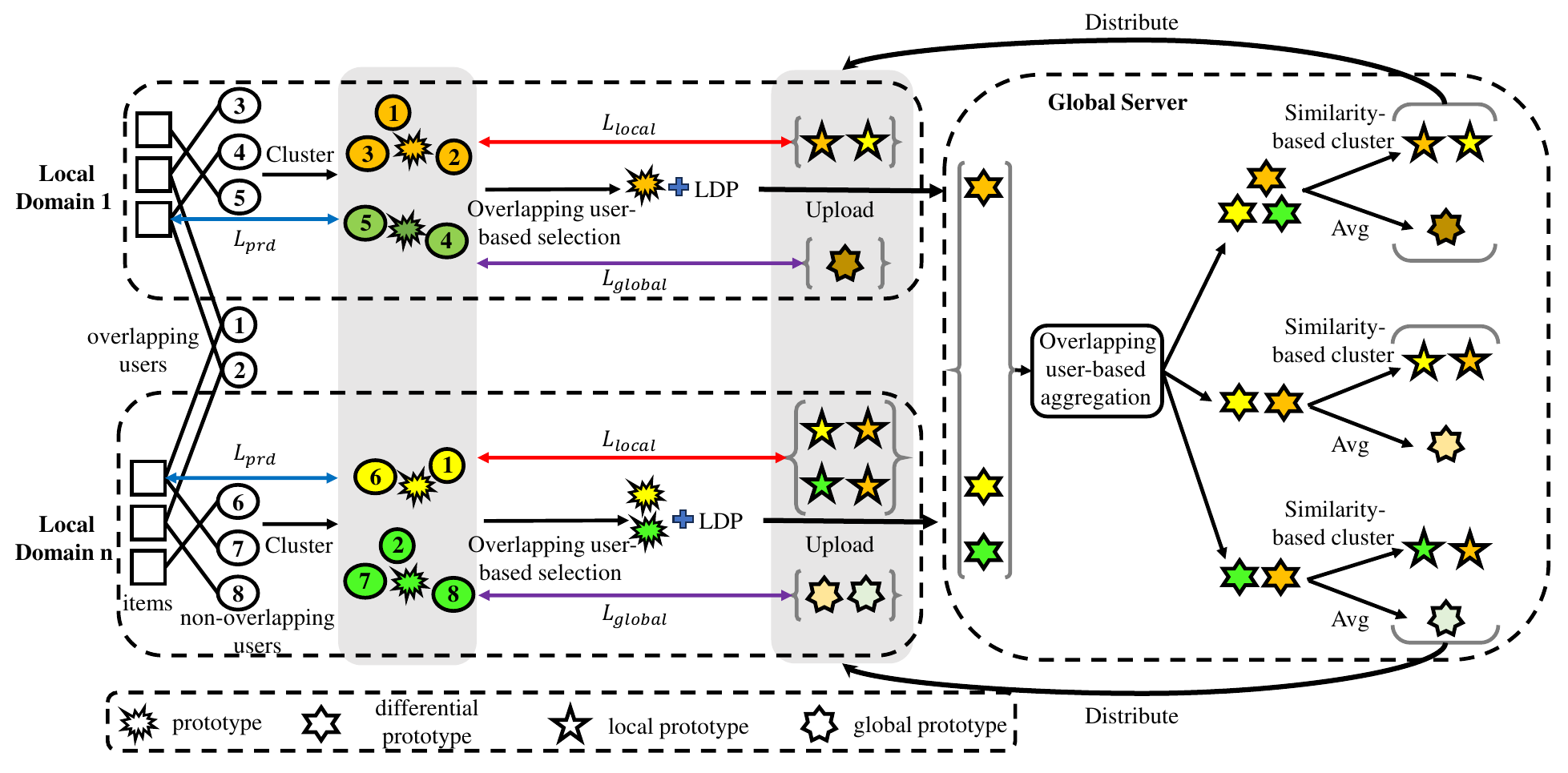}
\caption{The framework of FedPCL-CDR. (1) \textbf{Local Domain Learning}: (a) \textbf{Graph Representation Learning:} Introducing LightGCN to learn comprehensive user and item embeddings. 
(b) \textbf{Differential Prototype Learning:} Clustering all user embeddings and applying LDP to obtain differential prototypes.
(c) \textbf{Prototype-based Contrastive Learning:} Facilitating knowledge transfer across domains in a contrastive manner using both local and global prototypes. (2) \textbf{Global Server Aggregation}: It aggregates differential prototypes uploaded from multiple domains to generate local and global prototypes.}
\label{framework} %用于文内引用的标签
% \vspace{-1mm}
\end{figure*}

Figure \ref{framework} depicts the overall framework of FedPCL-CDR.
It mainly includes two key modules:
(1) \textbf{Local Domain Learning Module:} This module is designed to effectively utilize non-overlapping user information and transfer protected user interests across domains, which can be further divided into three components: (a) \textbf{Graph Representation Learning:} Introducing LightGCN to learn user and item embeddings. 
(b) \textbf{Differential  Prototype Learning:} Applying unsupervised k-means clustering to user embeddings to generate prototypes and then utilizing LDP to learn differential prototypes.
(c) \textbf{Prototype-based Contrastive Learning:} Facilitating knowledge transfer across domains in a CL manner using both local and global prototypes.
(2) \textbf{Global Server Aggregation Module:} This module focuses on modeling local and global prototypes by aggregating differential prototypes based on overlapping users.
\subsection{Local domain Learning Module}
\subsubsection{Graph Representation Learning}
Inspired by GNNs' ability to model complex user-item relationships, we use LightGCN \citep{he2020lightgcn} to learn embeddings for user and item IDs as well as their review texts. We construct a graph $G^i$, where nodes represent users and items, and edges indicate interactions between them. By utilizing the graph convolution and propagation layers of LightGCN, we encode user and item embeddings based on $G^i$. Specifically, we denote ID embeddings and review text embeddings at the $l$-th layer as $\textbf{E}_l^{i(id)}$ and $\textbf{E}_l^{i(rev)}$, respectively. Initially, ID embeddings $\textbf{E}_0^{i(id)}$ are randomly initialized, while review embeddings $\textbf{E}_0^{i(rev)}$ are learned using the document embedding model Doc2Vec \citep{le2014distributed}. The ID embeddings and review embeddings at the $l$-th layer are calculated as:
\begin{equation}
\begin{aligned}
\textbf{E}_l^{i(id)} &= (\textbf{D}^{-1/2}\textbf{A}\textbf{D}^{-1/2})\textbf{E}_{l-1}^{i(id)};\\
\textbf{E}_l^{i(rev)} &= (\textbf{D}^{-1/2}\textbf{A}\textbf{D}^{-1/2})\textbf{E}_{l-1}^{i(rev)},
 \end{aligned}
\end{equation}
where $\textbf{D}$ is a diagonal matrix and $\textbf{A}$ is an adjacency matrix.
After $l$ propagation steps, the final user and item ID embedding matrices  $\textbf{E}_u^{i(id)}$ and  $\textbf{E}_v^{i(id)}$ are generated by concatenating multiple embedding matrices from $\textbf{E}_0^{i(id)}$ to $\textbf{E}_l^{i(id)}$. Similarly, we obtain the user and item review text embedding matrix  
$\textbf{E}_u^{i(rev)}$ and $\textbf{E}_v^{i(rev)}$. Finally, we concatenate ID and review embeddings to learn comprehensive user and item embeddings:
% \begin{equation}
%  \begin{aligned}
% \textbf{E}_u^i &= f(\textbf{E}_u^{i(id)};\textbf{E}_u^{i(rev)});\\
% \textbf{E}_v^i &= f(\textbf{E}_v^{i(id)};\textbf{E}_v^{i(rev)}),
%  \end{aligned}
% \end{equation}
\begin{equation}
\textbf{E}_u^i = f_c(\textbf{E}_u^{i(id)};\textbf{E}_u^{i(rev)});
\textbf{E}_v^i = f_c(\textbf{E}_v^{i(id)};\textbf{E}_v^{i(rev)}),
\end{equation}
where $f_c$ represents the concatenation function. Here, we use the element-wise sum aggregation method.

\subsubsection{Differential Prototype Learning}
We begin by introducing the k-means algorithm \citep{macqueenclassification}, which aims to cluster user embeddings $\textbf{E}_u^i$ into $K$ groups. The cluster centroids serve as prototypes, representing the central characteristics of each user group with similar features. The selection of k-means clustering is motivated by its simplicity, efficiency, and proven effectiveness in grouping high-dimensional data. Unlike hierarchical or density-based methods, k-means scales well with large datasets and provides deterministic cluster assignments, which are crucial for generating stable prototypes. We obtain the prototype set:
\begin{equation}
T^i = \{\textbf{t}_j^i\}_{j=1}^K= Kmeans(\textbf{E}_u^i).
\label{kmeans}
\end{equation}

The clustering process incorporates data from all users within a domain, not limited to overlapping users.
This not only leverages the shared knowledge among overlapping users but also explores the effective 
utilization of knowledge from non-overlapping users.

We then select representative prototypes by considering overlapping users across domains. 
This involves choosing prototypes whose clusters include overlapping users. 
The rationale behind this choice lies in the intention that overlapping users have similar interests in different domains. 
The representative prototype set is calculated as follows:
\begin{equation}
C^i = \{\textbf{t}_j^i\}_{j=1}^K \xrightarrow{\text{overlapping users}} \{\textbf{c}_j^i\}_{j=1}^{K'},\ {K'}\le K.
\label{rep proto}
\end{equation}

These prototypes denote generalized user preferences, making it difficult for attackers to infer individual user details. To further improve privacy protection, we apply the LDP technique to the representative prototypes before transferring them. Specifically, we generate differential prototypes by adding Laplace noise to the representative prototypes:
\begin{equation}
\label{differential_proto}
    \hat{\textbf{c}}_j^i = clip(\textbf{c}_j^i,\beta)+Lap(0,\eta); \hat{C}^i=\{\hat{\textbf{c}}_j^i\}_{j=1}^{K'}, 
\end{equation}
where $\beta$ is the clipping threshold, and $\eta$ is the standard deviation of Laplace distribution. The LDP mechanism ensures that the leakage of private information is bounded by a privacy budget $\epsilon$, which has an upper bound $\frac{2\beta}{\eta}$, as derived in \citep{wu2022fedgnn}. A smaller privacy budget improves privacy protection but sacrifices the model's performance. In our experimental evaluation, we empirically examine the utility-privacy trade-off of LDP through the comprehensive analysis of the privacy budget.

In addition, we select overlapping users in each cluster to form the overlapping user set $O^i$:
\begin{equation}
O^i = \{o_j^i\}_{j=1}^{K'},\ o_j^i \subset U^o.
\label{overlap user}
\end{equation}

Finally, we upload each domain's differential prototype set $\hat{C}^i$ and overlapping user set $O^i$ to the global server. 

\subsubsection{Prototype-based Contrastive Learning} 
The server aggregates differential prototypes from all domains to generate both local and global prototypes (as detailed in the next section) and then distributes them back to the local domains.

To effectively transfer knowledge from other domains to the domain $D^i$, we enforce the alignment of user embeddings with corresponding global prototypes while distancing them from distinct global prototypes. The global prototype-based CL loss is defined as follows:
\begin{equation}
\small
L_{global}^i = -log\frac{exp(f_s(\textbf{e}_u^i,\textbf{g}_k^i))}{exp(f_s(\textbf{e}_u^i,\textbf{g}_k^i))+\sum_{\textbf{g}_j^i\in A(\textbf{g}_k^i),j\not = k}exp(f_s(\textbf{e}_u^i,\textbf{g}_j^i))},
\label{L_global}
\end{equation}
where $\textbf{g}_k^i$ denotes the global prototype corresponding to cluster $k$, to which the user embedding $\textbf{e}_u^i$ belongs. We regard $(\textbf{e}_u^i,\textbf{g}_k^i)$ that belongs to the same cluster as a positive pair.
Conversely, $\textbf{g}_j^i$ denotes the global prototype corresponding to cluster $j$, to which user embedding $\textbf{e}_u^i$ doesn't belong, and $(\textbf{e}_u^i,\textbf{g}_j^i)$ forms a negative pair.
$A(\textbf{g}_k^i)$ is the set of global prototypes excluding $\textbf{g}_k^i$.
$f_s$ indicates a similar function. We define it as:
\begin{equation}
f_s(\textbf{e}_u^i,\textbf{g}_k^i) = \frac{\textbf{e}_u^i\cdot \textbf{g}_k^i}{||\textbf{e}_u^i||\times||\textbf{g}_k^i||}/\tau,
\end{equation}
where $\tau$ represents the temperature coefficient, which controls the concentration strength of representation \citep{wang2021understanding}.

Furthermore, we introduce the local prototype-based CL loss to align $\textbf{e}_u^i$ with local prototypes of each domain through domain-wise CL in the latent space, enhancing inter-domain knowledge sharing. It is defined as:
% In addition to the global prototype-based contrastive learning loss, for the purpose of aligning $\textbf{e}_u^i$ with the local prototypes of each domain through alternating domain-wise contrastive learning in the latent space, facilitating greater inter-domain knowledge sharing, we introduce the local prototype-based contrastive learning loss as,

\begin{equation}
\small
L_{local}^i=-\frac{1}{M}\sum_{m=1}^Mlog\frac{exp(f(\textbf{e}_u^i,\textbf{l}_k^m))}{exp(f(\textbf{e}_u^i,\textbf{l}_k^m))+\sum_{\textbf{l}_j^i\in A(\textbf{l}_k^i),j\not=k}exp(f(\textbf{e}_u^i,\textbf{l}_j^i))},
\label{L_local}
\end{equation}
where $\textbf{l}_k^m$ denotes the local prototype of cluster $k$ from domains that have overlapping users with the cluster to which the user embedding $\textbf{e}_u^i$ belongs.
$\textbf{l}_k^i$ indicates the local prototype of cluster $k$ that includes $\textbf{e}_u^i$, and $A(\textbf{l}_k^i)$ is the set of local prototypes excluding $\textbf{l}_k^i$.

The global and local prototypes capture cluster-relevant information at different granularities, guiding the transfer of user interests from various perspectives.

\subsubsection{Local Training}
After refining the user embedding $\textbf{e}_u^i$, we concatenate it with the item embedding $\textbf{e}_v^i$ and process them through MLP layers for preference prediction.
The objective is to minimize the following loss:
\begin{equation}
L_{prd}^i = l(\hat{r}^i,r^i),
\label{L_prd}
\end{equation}
where $l$ is the cross-entropy loss function, $\hat{r}^i$ and $r^i$ are the predictive and ground-truth labels, respectively.
The total loss is:
\begin{equation}
L^i = L_{prd}^i + \alpha(L_{global}^i +L_{local}^i),
\label{loss function}
\end{equation}
where $\alpha$ is the trade-off parameter that balances $L_{prd}^i$ and prototype-based CL losses. The training procedure is detailed in Algorithm 1.

\begin{algorithm}[h!]
	%\textsl{}\setstretch{1.8}
\renewcommand{\algorithmicrequire}{\textbf{Input:}}
\renewcommand{\algorithmicensure}{\textbf{Output:}}
	\caption{FedPCL-CDR}
	\label{alg1}
	\begin{algorithmic}[1]
        \REQUIRE $D^i, \Theta_i, i=1,2,...,M$
        \STATE \textbf{Server executes:}
		\STATE Initialize global prototype set $G=\{G^1,G^2,...,G^M\}$ and local prototype set $L=\{L^1,L^2,...,L^M\}$.
		\FOR{round $r=1$ to $R$}
          \FOR{domain $i=1$ to $M$}
           \STATE $O^i,\hat{C}^i\gets LocalUpdate(i,G^i,L^i)$
          \ENDFOR
          \STATE Update global and local prototypes by Eq. (\ref{global_proto}) and Eq. (\ref{local proto}).
        \ENDFOR
        \STATE \textbf{LocalUpdate($i,G^i,L^i,$)}
		\FOR{epoch $e=1$ to $E$}
          \FOR{batch $b=1$ to $B$}
            \STATE Compute $L_{prd}^i$ by Eq. (\ref{L_prd}).
            \IF{\textbf{length}($G^i$)==0 and \textbf{length}($L^i$)==0}
                \STATE $L_{global}^i=0$, $L_{local}^i=0$
            \ELSE 
                \STATE Compute $L_{global}^i$ by Eq. (\ref{L_global}).
                \STATE Compute $L_{local}^i$ by Eq. (\ref{L_local}).
            \ENDIF
            \STATE Update $\Theta_i$ by Eq. (\ref{loss function}).
          \ENDFOR
        \ENDFOR
        \STATE Calculate prototype set $T^i$ by Eq. (\ref{kmeans}).
        \STATE Obtain representative prototype set $C^i$ by Eq. (\ref{rep proto}).
        \STATE Obtain differential prototype set $\hat{C}^i$ by Eq. (\ref{differential_proto}).
        \STATE Obtain overlapping user set $O^i$ by Eq. (\ref{overlap user}).
        \RETURN $O^i$ and $\hat{C}^i$
	\end{algorithmic}  
\end{algorithm}
\subsection{Global Server Aggregation}
After receiving differential prototype set $\hat{C}$ and overlapping user set $O$ from all domains, the global server calculates global prototypes for each domain.
First, for overlapping user set $o_k^i\in O^i$, we construct the prototype set $\bar{C}_k^i$ that includes all differential prototypes containing overlapping users with $o_k^i$ across domains:
\begin{equation}
\bar{C}_k^i = \bigcup_{{i'}\in D,{k'}\le{K'}}\{\hat{\textbf{c}}_{k'}^{i'}|o_{k'}^{i'}\cap o_k^i \not= \emptyset \}.
\end{equation}

Then, we calculate the global prototype $\textbf{g}_k^i$ for cluster $k$:
\begin{equation}
\label{global_proto}
\textbf{g}_k^i= \frac{1}{\hat{K}}\sum_{j=1}^{\hat{K}}\hat{\textbf{c}}_j^i, 
\end{equation}
where $\hat{K}$ is the length of prototype set $\bar{C}^i_k$.

Finally, we form the global prototype set $G^i=\{\textbf{g}_1^i, \textbf{g}_2^i,...,\textbf{g}_{K'}^i\}$.
The global prototype incorporates user preferences across domains from a high-level perspective.

Different from the global prototypes, for the local prototypes, we select some differential prototypes in all domains via similarity calculation.
Specifically, for each $\hat{\textbf{c}}_k^i$, we first calculate the cosine similarity between $\hat{\textbf{c}}_k^i$ and differential prototypes from other domains that include overlapping users with $o_k^i$. Then, we select the differential prototype with maximum similarity to $\textbf{c}_k^i$ in each domain to form the local prototype set $L_k^i$ for cluster $k$: 

\begin{equation}
L_k^i = \{\hat{\textbf{c}}_j^i\}_{j=1}^{\hat{K}} \xrightarrow{\text{similarity}} \{\textbf{l}_j^i\}_{j=1}^M, M\le \hat{K}.
\label{local proto}
\end{equation}

The local prototype set can be represented as $L^i=\{L_1^i,L_2^i,...,L_{K'}^i\}$.

After aggregation, the global server sends the global and local prototype sets $G^i$ and $L^i$ into the local domain $D^i$.

\subsection{Privacy Preserving Analysis}
The proposed FedPCL-CDR ensures user privacy through multiple mechanisms. Firstly, within the FL framework, data in each domain remains localized and is never shared with other domains, significantly reducing privacy risks \citep{wu2022fedgnn}. Secondly, cross-domain knowledge transfer is facilitated through prototypes, which inherently protect data privacy \citep{tan2022fedproto}. These prototypes are one-dimensional vectors derived from averaging low-dimensional representations of samples within the same group, making the process irreversible. Lastly, we incorporate LDP on prototypes before transferring them, ensuring that the privacy leakage is bounded and further strengthening privacy protection \citep{qi2020privacypreservinga}.

\subsection{Time Complexity Analysis} The total time complexity includes two main components: local domain learning and global server aggregation. For the local domain, the time complexity of graph learning is $O(|U^i|+
|V^i|+|Y|)dL$, where $|Y|$ is the number of user-item interactions, $d$ is the embedding dimension, and $L$ is the number of layers. The time complexity of k-means is $O(|U^i|KTd)$ where T is the number of iterations. The time complexity of prototype-based CL is $O(|U^i|(|L^i|+|G^i|)d)$. The time complexity of MLP layers is $O(|Y|dH)$ where $H$ is the number of MLP layers. For the global server aggregation, the time complexity is $O(K'{|\hat{C}_k^i|}^2+K'|\hat{C}_k^i|)$. Therefore, the overall time complexity is $M((|U^i|+
|V^i|+|Y|)dL+|U^i|KTd+|U^i|(|L^i|+|G^i|)d+|Y|dH)+K'{|\hat{C}_k^i|}^2+K'|\hat{C}_k^i|$. Due to $|U^i|+|V^i|<<|Y|$ and $|\hat{C}_k^i|<<|{\hat{C}_k^i|}^2$, The final time complexity can be simplified to  
$M(((L+H)|Y|+|U^i|(KT+|L^i|+|G^i|))d)+K'{|\hat{C}_k^i|}^2$.
\section{Experiments}
We evaluate FedPCL-CDR through extensive experiments across four CDR tasks constructed from the Amazon and Douban datasets. Our goal is to address the following research questions:
\begin{itemize}
    \item RQ1: How does FedPCL-CDR compare to state-of-the-art baselines in CDR tasks?
    \item RQ2: What is the contribution of each key component to FedPCL-CDR’s performance?
    \item RQ3: Does FedPCL-CDR address the sparse user overlap problem and ensure privacy protection? 
    \item RQ4: How sensitive is FedPCL-CDR to critical hyperparameters?
\end{itemize}

\subsection{Experimental Settings}
\subsubsection{Datasets}
Motivated by CDR methods \citep{chen2023win,zhang2023disentangled,cao2022disencdr,LIU2025107192}, we conduct experiments on four domains from the Amazon dataset\footnote{https://cseweb.ucsd.edu/~jmcauley/datasets/amazon/links.html}: Cell Phones and Accessories (Phone), Electronics (Elec), Clothing, Shoes and Jewelry (Cloth), and Sports and Outdoors (Sport), and three domains from the Douban dataset\footnote{https://www.dropbox.com/s/u2ejjezjk08lz1o/Douban.tar.gz?e=2\&dl=0}: Book, Movie, and Music.
These are organized into four CDR tasks. Basic dataset statistics are summarized in Table \ref{dataset_statistic}.
To align with implicit feedback settings, we binarize explicit ratings (0–5) into binary labels (1 or 0), consistent with the approach in \citep{ACHARYYA2025125667}. Following prior studies \citep{kang2019semi,zhao2020catn}, we apply standard filtering to enhance data quality, discarding users and items with fewer than 10 interactions.
\begin{table}[h]
\centering
\caption{Dataset statistics.}
\label{dataset_statistic}
% \fontsize{7.5}{8.5}\selectfont
% \setlength{\tabcolsep}{3.5pt}
\begin{tabular}{c|c|c|c|c|c|c}
\hline
Tasks&Datasets & \# Users & \# Items & \# \makecell{Overlapping\\ Users}& \# Interactions & Density\\
\hline
\multirow{2}*{Task 1}&Phone &5730	&22287 & \multirow{2}*{655} & 82111 & 0.064\%\\
&Sport &10849&35368 &  & 172241 & 0.045\%\\
\hline
\multirow{3}*{Task 2}&Elec &12301	&56081 &  \multirow{3}*{468} & 253300 & 0.037\%\\
&Cloth &13058&62137 &  & 185551 & 0.023\%\\
&Phone &5730	&22287 &  & 82111 & 0.064\%\\
\hline
\multirow{2}*{Task 3}&Movie &2320	&5803 & \multirow{2}*{1134} & 102864 & 0.764\%\\
&Music &1193&7146 & & 76592 & 0.898\%\\
\hline
\multirow{3}*{Task 4}&Book & 1715 & 8660 & \multirow{3}*{1008} & 104537 & 0.703\%\\
&Movie & 2320 &5803 & & 102864 & 0.764\%\\
&Music &1193	&7146 & &76592&0.898\%\\
\hline
\end{tabular}
\end{table}
\subsubsection{Evaluation Metrics}
Motivated by previous CDR practice \citep{liu2020cross,zhu2023domain}, we evaluate model performance via the leave-one-out approach. 
For each user, one interaction is randomly selected to form the test set, while the rest are used for training. 
Following  \citep{he2017neural,wang2025causal}, we sample 99 uninteracted items as negative samples for each test user, with the actual user-item interaction serving as the positive sample. The FedPCL-CDR model then generates prediction scores for all 100 candidates to facilitate ranking. 
Evaluation is conducted using Hit Ratio (HR) and Normalized Discounted Cumulative Gain (NDCG), which are frequently used in CDR methods \citep{li2020ddtcdr,zhu2019dtcdr,zhu2021unified,NI2024106488}.

\subsubsection{Parameter Settings}
 We obtain optimal hyperparameters by optimizing the loss function \eqref{loss function} using the Adam optimizer. The learning rate is set to 0.001. 
 The weight $\alpha$ for the prototype-based CL losses is set to 0.01. Additionally, embedding vectors are 64-dimensional, and training is performed with a batch size of 256. The temperature coefficient in CL is established at 0.2, and the cluster number is set to 10. Furthermore, we apply batch normalization, dropout, and early stopping techniques to prevent overfitting. All baselines are executed using their GitHub source code, with hyperparameters carefully tuned on our datasets for the best performance. Experiments are conducted using two NVIDIA Testra V100 (32G) GPUs.

 \subsubsection{Baseline Methods.}
%To validate the effectiveness of our model, 
We evaluate FedPCL-CDR against several SOTA baselines, which are commonly adopted in recent CDR research \citep{lu2023contrastive, lin2024enhancing}.
\begin{itemize}
    \item \textbf{NeuMF} \citep{he2017neural} fuses collaborative filtering with neural networks to model user-item interactions. 
    \item \textbf{LightGCN} \citep{he2020lightgcn} simplifies GCN by propagating user and item embeddings directly through the interaction graph, avoiding complex operations or auxiliary data. 
    \item \textbf{FedNCF} \citep{perifanis2022federated} introduces FL into the model Neural Collaborative Learning (NCF) to protect user's privacy. 
\item \textbf{GA-DTCDR} \citep{zhu2020graphical} proposes a graphical and attentional framework to learn comprehensive user and item embeddings. 
\item \textbf{NMCDR} \citep{xu2023neural} is a neural node matching framework that conducts intra-knowledge and inter-knowledge fusion to improve the model performance. 
\item \textbf{CL-DTCDR} \citep{lu2023contrastive} introduces both intra-domain and inter-domain CL tasks to address the data-sparsity issue and facilitate efficient knowledge transfer.
\item \textbf{GA-MTCDR-P} \citep{zhu2021unified} is an enhanced model of GA-DTCDR, which incorporates graph neural networks and multi-domain attention mechanisms to address the problem of negative transfer in multi-target domain recommendations.
 \item \textbf{PriCDR} \citep{chen2022differential} is a PPCDR framework that leverages DP to release the source domain's rating matrix, which is subsequently transferred to the target domain. 
 \item \textbf{P2FCDR} \citep{chen2023win} is a federated CDR model that incorporates LDP to safeguard user embeddings during inter-domain knowledge transfer.
 \item ${\textbf{P}^\textbf{2}}$\textbf{DTR} \citep{chen2023winwin} introduces a dual-target CDR framework that utilizes a private set intersection algorithm and prototype-based FL to protect user privacy.
 \end{itemize}

 \begin{table*}[t!]
\centering
\caption{Experimental results on four CDR tasks. The best performance is in bold, and the second best is underlined. All improvements are significant over baselines (t-test with $p < 0.01$).}
\fontsize{7.5}{8.5}\selectfont
\setlength{\tabcolsep}{3.0pt}
\begin{tabular}{ c| c| c c c c c c c c c c c c}
\hline
Domain & Metric & NeuMF & LightGCN & FedNCF & \makecell{GA-\\DTCDR} &\makecell{GA-\\MTCDR-P}&NMCDR &\makecell{CL-\\DTCDR}&PriCDR&P2FCDR&$P^2DTR$&\makecell{FedPCL\\-CDR}&Imp\\
\hline
\multirow{2}*{Phone}&HR& 0.3490&0.3552&0.3477&0.3809&0.3827&0.3841&\underline{0.5277}&0.3553&0.3756&0.3843&\textbf{0.5501}&2.24\%\\
& NDCG& 0.2141&0.2267&0.2034&0.2285&0.2304&0.2349&\underline{0.3193}&0.2259&0.2267&0.2401&\textbf{0.3541}&3.48\%\\
\multirow{2}*{Sport}&HR& 0.3134&0.3257&0.3018&0.3749&0.3799&0.3914&\underline{0.5660}&0.3687&0.4235&0.4326&\textbf{0.5991}&3.31\%\\
 & NDCG & 0.1886&0.1939&0.1593&0.2056&0.2148&0.2166&\underline{0.3344}&0.2389&0.2578&0.2645&\textbf{0.3812}&4.68\%\\
\hline
\multirow{2}*{Elec}&HR&0.4065&0.4124&0.3921&0.4568&0.4585&\underline{0.5234}&0.5204&0.4234&0.4873&0.4987&\textbf{0.5408}&1.74\%\\
 & NDCG & 0.3000&0.3015&0.2923&0.3102&0.3117&0.3016&0.3292&0.3234&0.3252&\underline{0.3325}&\textbf{0.3445}&1.20\%
\\
\multirow{2}*{Cloth}&HR& 0.2207&0.2367&0.2133&0.3044&0.2969&0.3030&\underline{0.4235}&0.2816&0.3021&0.3124&\textbf{0.4489}&2.54\%\\
 & NDCG & 0.1215&0.1402&0.1045&0.2156&0.2024&0.1965&\underline{0.2511}&0.1856&0.2035&0.2201&\textbf{0.2678}&1.67\%\\
\multirow{2}*{Phone}&HR&0.3490&0.3552&0.3477&0.4021&0.4093&0.4823&\underline{0.5159}&0.3915&0.4329&0.4451&\textbf{0.5341}&1.82\%\\
 & NDCG & 0.2141&0.2267&0.2034&0.2614&0.2705&\underline{0.3142}&0.3059&0.2621&0.2804&0.2943&\textbf{0.3417}&2.75\%\\
\hline
\multirow{2}*{Movie}&HR& 0.3294&0.3325&0.3145&0.3629&0.3691&0.3787&0.3525&0.3459&0.3703&\underline{0.3825}&\textbf{0.4336}&5.11\%\\
 & NDCG&0.1768&0.1867&0.1505&0.1955&0.2014&0.2156&0.2016&0.2186&0.2037&\underline{0.2201}&\textbf{0.2301}&1.00\%\\
\multirow{2}*{Music}&HR&0.2686&0.2639&0.2987&0.3142&\underline{0.3187}&0.3025&0.3151&0.3044&0.3023&0.3156&\textbf{0.3294}&1.07\%\\
 & NDCG & 0.1468&0.1503&0.1589&0.1680&0.1704&\underline{0.1788}&0.1631&0.1772&0.1659&0.1703&\textbf{0.1842}&0.54\%\\
\hline 
\multirow{2}*{Book}&HR& 0.2907&0.3045&0.1847&0.3069&0.3091&0.3096&\underline{0.3431}&0.2525&0.2684&0.2764&\textbf{0.3727}&2.96\%\\
 & NDCG& 0.1830&0.1904&0.1409&0.1826&0.1918&0.2016&\underline{0.2072}&0.1902&0.1932&0.2031&\textbf{0.2265}&1.93\%\\
\multirow{2}*{Movie}&HR& 0.3294&0.3325&0.3145&0.3740&0.3687&0.3703&\underline{0.4328}&0.3336&0.3856&0.3902&\textbf{0.4470}&1.42\%\\
 & NDCG & 0.1768&0.1867&0.1505&0.1894&0.1769&0.1968&\underline{0.2221}&0.1530&0.2046&0.2135&\textbf{0.2334}&1.13\%\\
\multirow{2}*{Music}&HR& 0.2686&0.2639&0.2987&0.3142&0.3178&\underline{0.3225}&0.3121&0.3050&0.3069&0.3124&\textbf{0.3418}&1.93\%\\
 & NDCG & 0.1468&0.1503&0.1689&0.1780&0.1795&\underline{0.1888}&0.1877&0.1685&0.1736&0.1735&\textbf{0.2034}&0.47\%\\
\hline 
\end{tabular}
%\vspace{-4mm}
\label{comparison}
\end{table*}
\subsection{Experimental Results and Analysis}
 \subsubsection{Performance Evaluation (RQ1)}
FedPCL-CDR and baselines are evaluated using standard metrics including HR@10 and NDCG@10. From the experimental results in Table \ref{comparison}, We can observe that:  \begin{itemize}
 \item Our model, FedPCL-CDR, surpasses other baselines, showcasing its capability to achieve satisfactory performance while also safeguarding user privacy. Specifically, FedPCL-CDR outperforms the top-performing CDR baseline by an average of 2.45\% in HR@10 and 1.93\% in NDCG@10 across all tasks. This improvement can be attributed to the following reasons: \textbf{(1)} 
 FedPCL-CDR efficiently utilizes non-overlapping user data to transfer cross-domain knowledge, which is particularly beneficial in scenarios with sparse overlapping users, such as Phone\&Sport. \textbf{(2)} By constructing dual prototype-based CL tasks, FedPCL-CDR achieves more effective knowledge transfer.
  \item FedPCL-CDR outperforms single-domain federated methods such as FedNCF. This demonstrates the significant role of cross-domain knowledge in enhancing recommendation performance within the FL framework. Furthermore, FedPCL-CDR outperforms the PPCDR methods PriCDR and P2FCDR, demonstrating that leveraging prototypes for knowledge transfer not only protects user privacy but also improves model performance. Moreover, FedPCL-CDR performs better than $P^2$DTR, demonstrating that utilizing non-overlapping user information and dual-prototype CL learning paradigm can improve the recommendation performance. %Moreover, FedPCL-CDR exceeds the performance of GA-DTCDR, which depends on fully overlapping users for knowledge transfer. This indicates that effectively utilizing non-overlapping user information can improve model performance. 
  Finally, although our method and CL-DTCDR both use CL to transfer knowledge, FedPCL-CDR still performs better than CL-DTCDR, showing that our method not only protects user privacy but also improves model performance.
 \item  GNN-based methods (e.g., LightGCN) surpass non-graph approaches (e.g., NeuMF), which shows that leveraging high-order neighborhood information improves accuracy.
 \item CDR methods consistently outperform single-domain approaches, as evidenced by the comparison between GA-DTCDR and NeuMF. This shows that cross-domain knowledge can alleviate the data-sparsity issue.
 % \item  Single-target CDR methods generally exhibit inferior performance compared to multi-target CDR methods, such as PriCDR vs NMCDR. This discrepancy can be attributed to the fact that multi-target methods can simultaneously enhance the performance of multiple domains.
 \end{itemize}

\subsubsection{Ablation Studies (RQ2)}
To evaluate FedPCL-CDR's components, we created three variants: 
%of FedPCL-CDR, denoted as w/o glob-proto, w/o loc-proto, and w/o rev by removing specific components.
\textbf{(1)} \textbf{w/o loc-proto}: we eliminate the local prototype-based CL loss.
\textbf{(2)} \textbf{w/o glob-proto}: we remove the global prototype-based CL loss. \textbf{(3) w/o LDP}: We remove the LDP mechanism during FedPCL-CDR training, enabling direct prototype transfer across domains without noise injection.
Table \ref{ablation} presents the results. We can observe that: 
\textbf{(1)} The performance drop in models \textbf{w/o loc-proto} and \textbf{w/o glob-proto} highlights the crucial role of both local and global prototype-based CL in achieving superior performance. 
\textbf{(2)} In general, \textbf{w/o loc-proto} contributes more, which shows that local prototype-based CL plays an important role in improving model performance. \textbf{(3)} While \textbf{w/o LDP} yields higher accuracy by omitting noise injection, this variant increases the risk of privacy leakage.
In conclusion, each component in FedPCL-CDR plays a crucial role, demonstrating the rationality and effectiveness of our design. 
\begin{table}[h!]
\centering
\caption{Ablation studies on FedPCL-CDR.}
% \fontsize{8.0}{8.5}\selectfont
% \setlength{\tabcolsep}{3.5pt}
\begin{tabular}{c| c| c| c c c c}
\hline
Task & Domain & Metric  & \makecell{w/o \\loc-proto} & \makecell{w/o \\glob-proto}& \makecell{w/o \\ LDP}&FedPCL-CDR\\
\hline
\multirow{4}*{Task 1}&\multirow{2}*{Phone}&HR& 0.5478&0.5364&0.5933&0.5501 \\
& & NDCG& 0.3532&0.3487&0.3949&0.3541\\
&\multirow{2}*{Sport}&HR& 0.5868&0.5938&0.6766&0.5991 \\
& & NDCG & 0.3760&0.3788&0.4534&0.3812 \\
\hline 
\multirow{6}*{Task 2}&\multirow{2}*{Elec}&HR& 0.5371&0.5248&0.6078&0.5408 \\
& & NDCG&0.3413&0.3356&0.4018&0.3445\\
&\multirow{2}*{Cloth}&HR&0.4324&0.4455&0.5018&0.4489\\
& & NDCG &  0.2641&0.2590&0.3166&0.2678\\
&\multirow{2}*{Phone}&HR&0.5222&0.5240&0.5970&0.5341  \\
& & NDCG &  0.3304&0.3312&0.3844&0.3417\\
\hline 
\multirow{4}*{Task 3}&
\multirow{2}*{Movie}&HR& 0.4232&0.4241&0.4379&0.4336 \\
& & NDCG&0.2291&0.2293&0.2356&0.2301 \\
& \multirow{2}*{Music}&HR&0.3203&0.3194&0.4518&0.3294  \\
& & NDCG& 0.1801&0.1736&0.2487&0.1842\\
\hline
\multirow{6}*{Task 4}&\multirow{2}*{Book}&HR& 0.3656&0.3662&0.4724&0.3727   \\
& & NDCG&  0.2171&0.2058&0.2865&0.2265  \\
&\multirow{2}*{Movie}&HR& 0.4414&0.4444&0.4610&0.4470 \\
& & NDCG & 0.2317&0.2283&0.2444&0.2334 \\
&\multirow{2}*{Music}&HR& 0.3401&0.3385&0.4745&0.3418 \\
& & NDCG & 0.1907&0.1860&0.2059&0.1935 \\
\hline 
\end{tabular}
%\vspace{-4mm}
\label{ablation}
\end{table}

\subsubsection{Performance for different proportions of overlapping users (RQ3)}
To assess FedPCL-CDR's capability in addressing sparse overlapping users within CDR, we manipulate the overlapping ratio specifically for Task 1 across different settings. These varying ratios signify different levels of commonality, where a higher ratio indicates a greater number of overlapping users across domains. For instance, in Task 1 with the ``Phone-Sport" dataset and an overlapping ratio of 30\%, the number of overlapping users is computed as 655 * 30\% = 196. We report results about several representative CDR methods. The corresponding results with different overlapping ratios are shown in Table \ref{user_ratio}. As the overlapping ratio increases, the performance of all models demonstrates improvement. This is intuitively sensible, as a higher overlapping ratio implies a greater number of shared users, facilitating a more straightforward transfer of cross-domain knowledge. The performance of GA-DTCDR, PriCDR, and $P^2$DTR shows significant fluctuations, primarily because they rely on overlapping users to transfer knowledge. In contrast, both NMCDR, CL-DTCDR, and our method, FedPCL-CDR, demonstrate relatively minor changes, indicating that effectively transferring knowledge across non-overlapping users can enhance performance and ensure model stability.
 \begin{table}[!htb]
\centering
\caption{Experimental results on Task 1 with different overlapping user ratios.}
% \fontsize{7}{7.0}\selectfont
% \setlength{\tabcolsep}{2.5pt}
\begin{tabular}{c| c| c| c c c c c c}
\hline
Domain&\makecell{overlapping\\ user ratio}&Metric&GA-DTCDR&NMCDR&PriCDR&\makecell{CL\\-DTCDR}&$P^2$DTR&\makecell{PedPCL\\ -CDR}\\
\hline
\multirow{8}*{Phone}&\multirow{2}*{30\%}&HR&0.2042&0.3649& 0.2041&0.4954&0.3014&0.5223 \\
& & NDCG&0.1022&0.1994&0.1053 &0.2685&0.1874&0.3368 \\
& \multirow{2}*{50\%}&HR&0.2691&0.3762&0.2110 &0.5047&0.3204&0.5353\\
& & NDCG&0.1426&0.2078&0.1185 &0.2736&0.2018&0.3402 \\
& \multirow{2}*{70\%}&HR&0.2838&0.3733&0.2449 &0.5124&0.3578&0.5412 \\
& & NDCG&0.1596&0.2037&0.1464 &0.2731&0.2201&0.3421\\
%& \multirow{2}*{100\%}&HR&0.3809&0.3841&0.3753 &0.5277&0.5933\\
%& & NDCG&0.2285&0.2349&0.2259 &0.3193&0.3949\\
\hline
\multirow{8}*{Sport}&\multirow{2}*{30\%}&HR&0.2520&0.3746&0.1429 &0.5292&0.3604&0.5649\\
& & NDCG&0.1413&0.1951&0.0743 &0.3173&0.1903&0.3557 \\
& \multirow{2}*{50\%}&HR&0.2910&0.3863&0.1865 &0.5381&0.3892&0.5781 \\
& & NDCG&0.1713&0.2026&0.1002 &0.3203&0.2213&0.3679\\
& \multirow{2}*{70\%}&HR&0.3144&0.3908&0.2271 &0.5493&0.4110&0.5845\\
& & NDCG&0.1983&0.2128&0.1286 &0.3302&0.2403&0.3746 \\
%& \multirow{2}*{100\%}&HR&0.3749&0.3914&0.4087 &0.5660&0.6766\\
%& & NDCG&0.2056&0.2166&0.2389 &0.3344&0.4534 \\
\hline
\end{tabular}
\label{user_ratio}
% \vspace{-1mm}
\end{table}

\subsubsection{Empirical Study of Privacy (RQ3)}
We compare the privacy-preserving capabilities of FedPCL-CDR and PPCDR baselines by simulating an attack where an attacker attempts to reconstruct original user embeddings. Assuming the attacker intercepts user features during client-server communication, they use a deep neural network to infer the original embeddings, minimizing the reconstruction error.  
We use Mean Squared Error (MSE) as an evaluation metric to measure the accuracy of the reconstructed embeddings compared to the original ones. A higher MSE value indicates greater resistance to original embedding reconstruction, demonstrating stronger privacy preservation. We conducted experiments on Tasks 1 and 3 and reported the results in Table \ref{privacy_test}. We can find that FedPCL-CDR obtains the optimal privacy-preserving capabilities.

\begin{figure}[!htbp]
	\centering
 \captionsetup[subfigure]{} 
	\subfloat[Phone\&Sport]{\includegraphics[width=.4\linewidth,height=0.25\textheight]{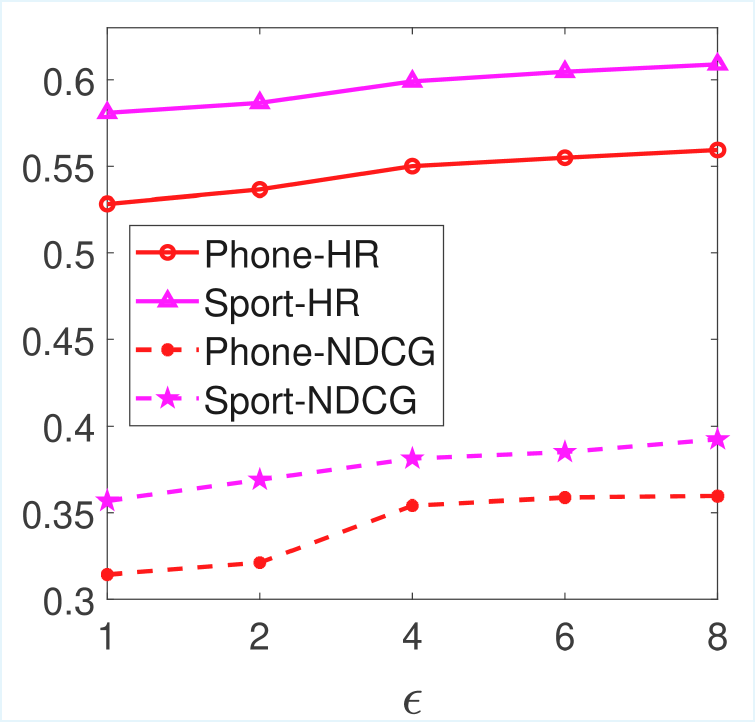}}\hspace{3pt}
	\subfloat[Movie\&Music]{\includegraphics[width=.4\linewidth,height=0.25\textheight]{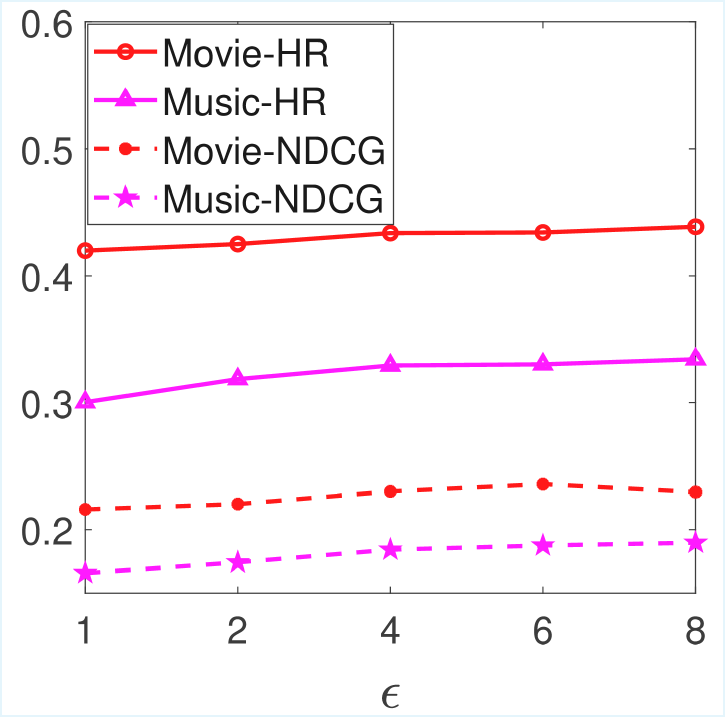}}\hspace{3pt}
	\caption{Performance of different privacy budgets on Tasks 1 and 3.}
    \label{privacy_budget}
    % \vspace{-4mm}
\end{figure}

Additionally, we analyze the trade-off between recommendation performance and privacy, as shown in Figure \ref{privacy_budget}. A larger privacy budget $\epsilon$ results in improved performance but leads to greater privacy leakage. Notably, Notably, performance variation becomes relatively stable when $\epsilon>=4$, indicating this threshold represents an optimal privacy-accuracy trade-off. Therefore, we set $\epsilon=4$ to achieve a balance between performance and privacy.

\begin{table}[h!]
% \vspace{-1mm}
\centering
\caption{MSE on Tasks 1 and 3.}
% \fontsize{7}{7.5}\selectfont
% \setlength{\tabcolsep}{4pt}
\begin{tabular}{c| c| c| c c c}
\hline
Task & Domain & Metric  & P2FCDR & $P^2DTR$& FedPCL-CDR\\
\hline
\multirow{2}*{Task 1}&Phone&MSE&5.12&6.74&8.72\\
&Sport&MSE&9.25&10.25&12.67\\
\hline 
\multirow{2}*{Task 3}&Movie&MSE&3.21&3.78&4.48\\
&Music&MSE&3.46&4.32&5.67\\
\hline 
\end{tabular}
%\vspace{-4mm}
\label{privacy_test}
\end{table}

\subsubsection{Impact of Hyper-parameters (RQ4)}
We assess FedPCL-CDR's performance across different configurations of three pivotal parameters: the weight parameter $\alpha$ associated with prototype-based CL losses, the cluster number $K$, and the recommendation list length $N$. 
%We report HR@10 and NDCG@10 on Tasks 1 (Phone\&Sport) and 3 (Movie\&Music).
\begin{itemize}
    \item \textbf{Impact of $\alpha$.}
 We employ the parameter $\alpha$ to control the degree of knowledge transfer across domains. To evaluate its impact, we conduct experiments with different $\alpha$ values, namely $[0.001, 0.01, 0.1, 0.2]$. Figure \ref{alpha} illustrates the outcomes for HR@10 and NDCG@10.
We find that as the weight parameter $\alpha$ increases, the performance first rises and then decreases. The best performance of FedPCL-CDR is observed at $\alpha = 0.01$. 
\begin{figure}[htbp]
	\centering
 \captionsetup[subfigure]{}
	\subfloat[Phone\&Sport]{\includegraphics[width=.4\linewidth,height=0.25\textheight]{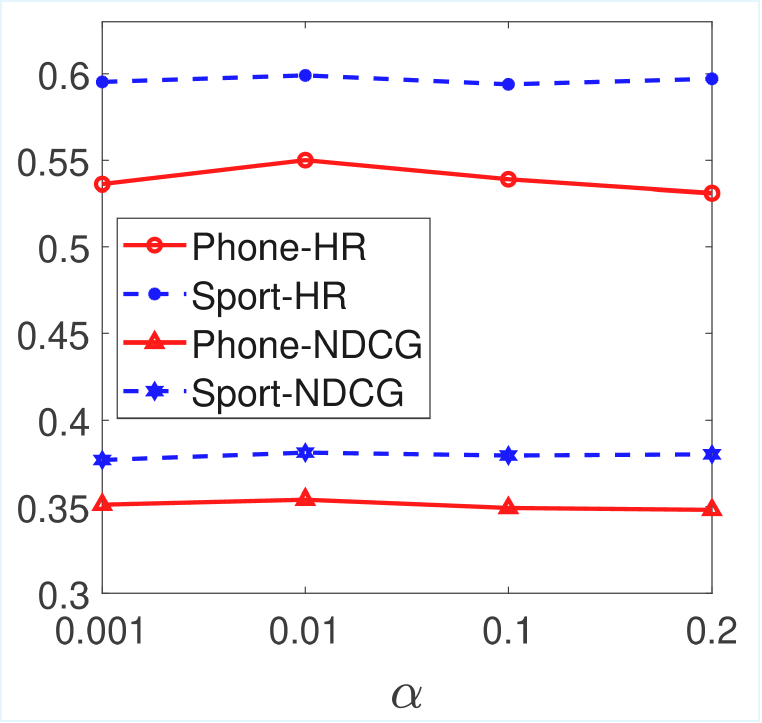}}\hspace{3pt}
	\subfloat[Movie\&Music]{\includegraphics[width=.4\linewidth,height=0.25\textheight]{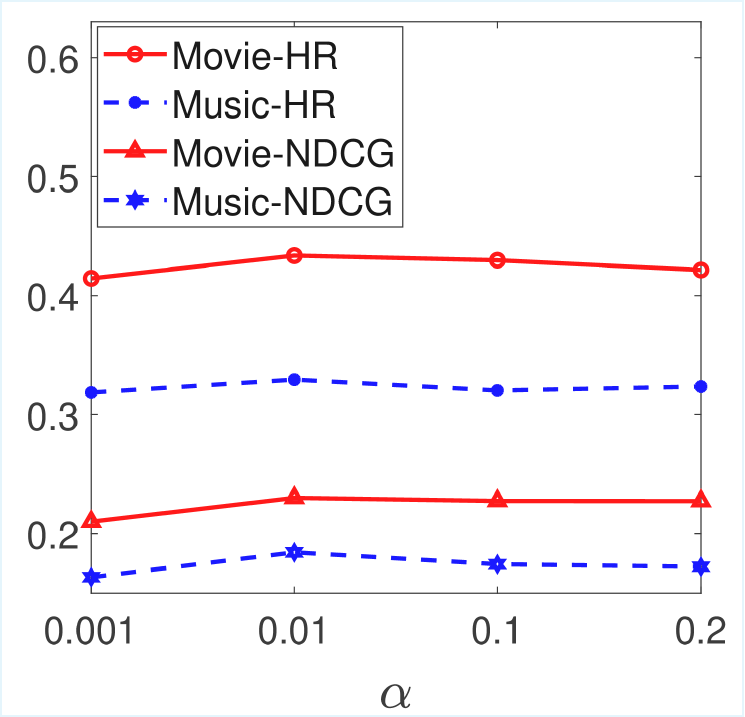}}\hspace{3pt}
	\caption{Performance of different $\alpha$.}
  \label{alpha}
  %\vspace{-4mm}
\end{figure}
\item \textbf{Impact of cluster number $K$.}
The number of clusters significantly influences the generalization of prototypes, thereby affecting the learning of user preferences. As shown in Figure \ref{cluster_num}, FedPCL-CDR reaches its peak performance when the number of clusters is set to 10. With the increase in the number of clusters, the HR@10 and NDCG@10 metrics initially rise, reaching a maximum at $K=10$, and subsequently decline. This trend can be attributed to the fact that an excessive number of clusters results in overly specific prototypes, which lack generalization capabilities and lead to suboptimal knowledge transfer.
\begin{figure}[!htbp]
	\centering
 \captionsetup[subfigure]{}
	\subfloat[Phone\&Sport]{\includegraphics[width=.4\linewidth,height=0.25\textheight]{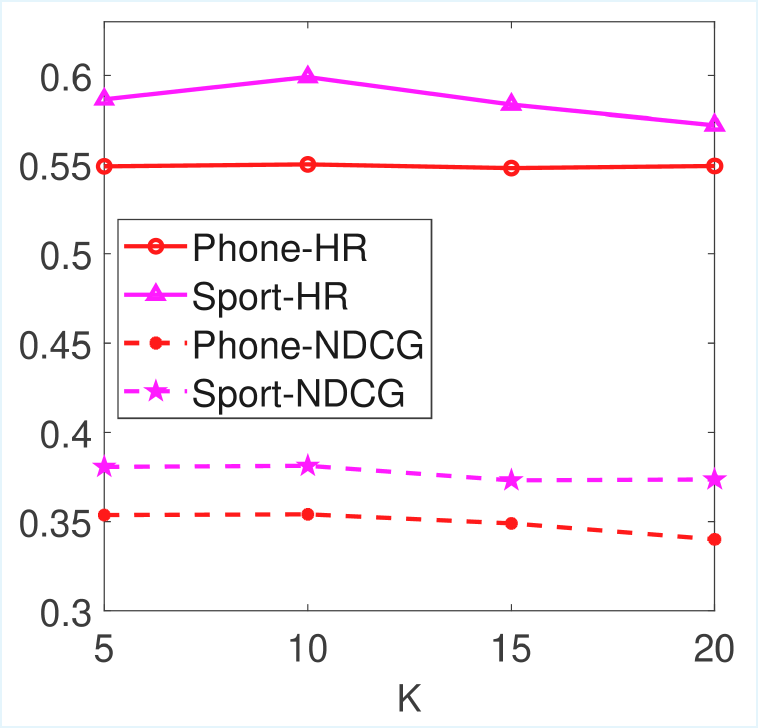}}\hspace{3pt}
	\subfloat[Movie\&Music]{\includegraphics[width=.4\linewidth,height=0.25\textheight]{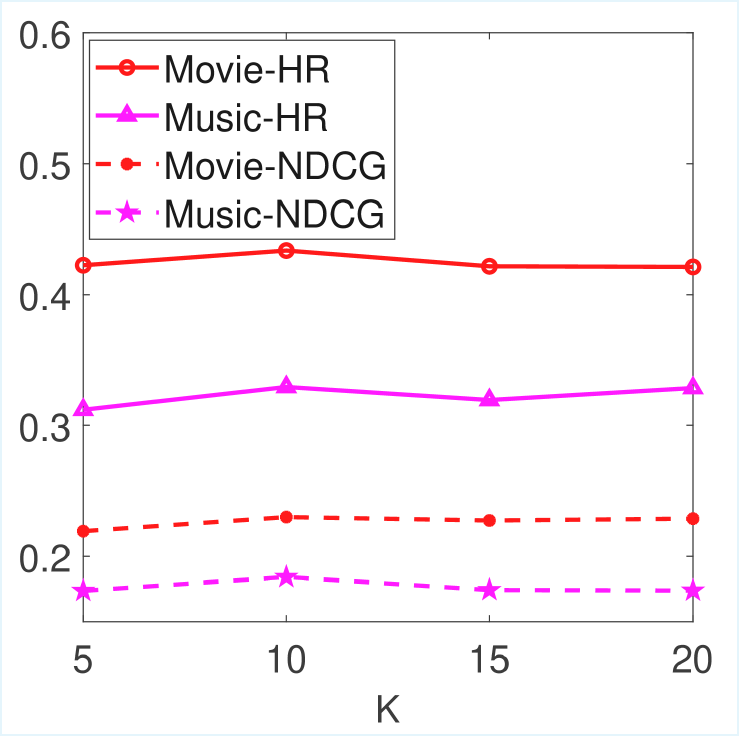}}\
	\caption{Performance of different cluster number K.}
  \label{cluster_num}
   %\vspace{-4mm}
\end{figure}

\item \textbf{Impact of recommendation list length $N$.}
To assess how the length of the recommendation list influences performance, we evaluate FedPCL-CDR under different values of $N \in \{2, 4, 6, 8, 10\}$, as shown in Figure~\ref{top_N}. The results reveal a consistent performance improvement as $N$ increases. This trend is expected, since recommending a greater number of items increases the likelihood of including relevant ones, thereby simplifying the task and boosting model effectiveness.
\begin{figure}[htbp]
	\centering
 \captionsetup[subfigure]{}
	\subfloat[Phone\&Sport]{\includegraphics[width=.4\linewidth,height=0.25\textheight]{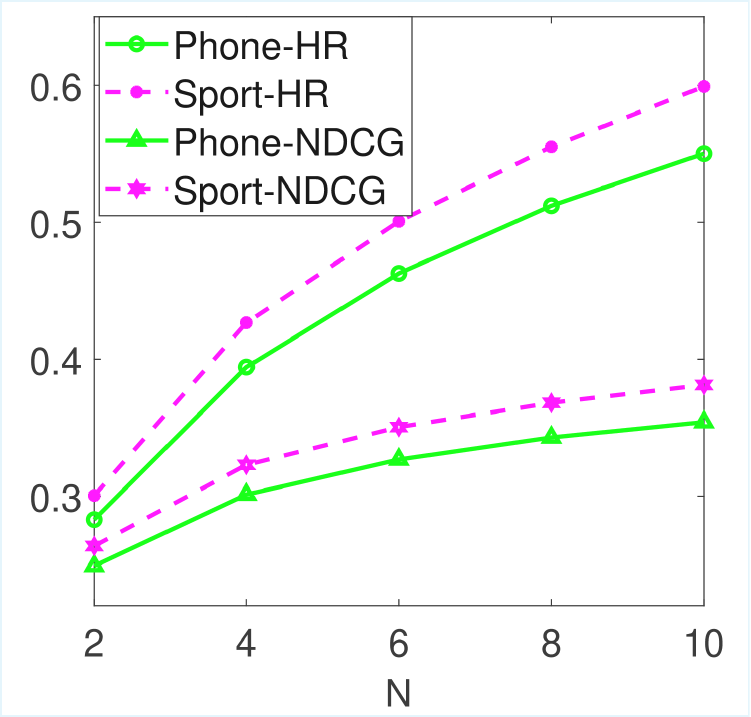}}\hspace{3pt}
	\subfloat[Movie\&Music]{\includegraphics[width=.4\linewidth,height=0.25\textheight]{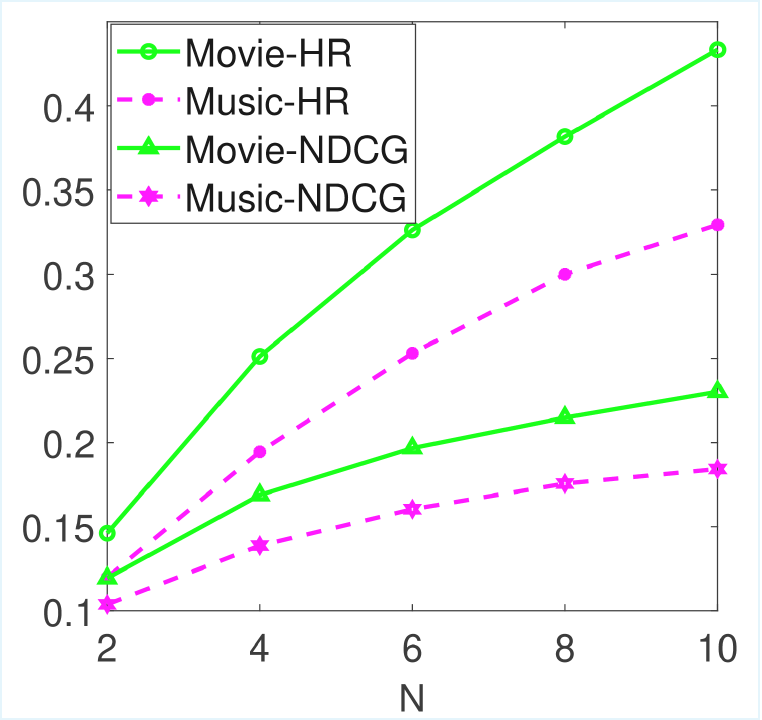}}\
	\caption{Performance of different recommendation list length N.}
  \label{top_N}
%  \vspace{-4mm}
\end{figure}
\end{itemize}

\section{Conclusion and Future work}
In this paper, we introduce FedPCL-CDR, a federated prototype-based CL framework for PPCDR. It aims to address the sparse overlapping problem and enhance user privacy preservation. FedPCL-CDR comprises two modules: (1) a local domain learning module that privately transfers knowledge of both overlapping and non-overlapping users across domains using differential prototypes in a CL manner, and (2) a global server aggregation module that aggregates user interests from multiple domains by modeling local and global prototypes. We have conducted extensive experiments on four CDR tasks using two real-world datasets to demonstrate the effectiveness of our proposed model. 

While our proposed framework, FedPCL-CDR, effectively addresses the sparse overlapping user problem while enhancing privacy protection, it has several limitations. First, although prototype transmission reduces communication costs, frequent server-client synchronization in federated learning may still introduce latency in cross-device deployments. Second, while our method can operate with very sparse overlaps, it fundamentally requires some overlapping users to select representative prototypes, making fully non-overlapping scenarios unsolved. Finally, the fixed cluster number may not adapt well to domains with highly dynamic user interests. We plan to address these limitations in future work.

\section*{Acknowledgements}
This work is supported by the Australian Research Council (LP210100129).

\section*{CRediT authorship contribution statement}
\textbf{Li Wang}: Conceptualization, Investigation, Methodology,
Software, Writing - original draft, Writing - review \& editing.
\textbf{Qiang Wu}: Supervision, Writing - review \& editing.
\textbf{Min Xu}: Supervision, Writing - review \& editing.

% \appendix
% \section{Example Appendix Section}


% %% For citations use: 
% %%       \citep{<label>} ==> [1]

% %%
% Example citation, See \citep{lamport94}.

 \bibliographystyle{elsarticle-harv} 
 \bibliography{ref.bib}




% %% For numbered reference style

%   \textit{\LaTeX: a document preparation system},
%   Addison Wesley, Massachusetts,

% \end{thebibliography}
\end{document}